\documentclass{ws-m3as}
\usepackage{amssymb}
\usepackage{amsmath,amssymb}
\usepackage{latexsym} 
\usepackage{graphicx}
\usepackage{subfigure}
\usepackage{float}

\begin{document}

\markboth{Hui-Hui Dai and Zilong Song}{Analytical Solutions for the
Equilibrium states}

\catchline{}{}{}{}{}

\title{Analytical Solutions for the Equilibrium states of a Swollen Hydrogel
Shell and an Extended Method of Matched Asymptotics}

\author{Hui-Hui Dai and Zilong Song}
\address{Department of Mathematics, City University of Hong Kong,\\
83 Tat Chee Avenue, Kowloon Tong, Hong Kong\\
mahhdai@cityu.edu.hk}

\maketitle

\begin{abstract}
A polymer network can imbibe water, forming an aggregate called
hydrogel, and undergo large and inhomogeneous deformation with
external mechanical constraint. Due to the large deformation,
nonlinearity plays a crucial role, which also causes the mathematical
difficulty for obtaining analytical solutions. Based on an existing
model for equilibrium states of a swollen hydrogel with a core-shell
structure, this paper seeks analytical solutions of the deformations
by perturbation methods for three cases, i.e. free-swelling, nearly
free-swelling and general inhomogeneous swelling. Particularly for the
general inhomogeneous swelling, we introduce an extended method of
matched asymptotics to construct the analytical solution of the
governing nonlinear second-order variable-coefficient differential
equation. The analytical solution captures the boundary layer behavior
of the deformation. Also, analytical formulas for the radial and hoop
stretches and stresses are obtained at the two boundary surfaces of
the shell, making the influence of the parameters explicit. An
interesting finding is that the deformation is characterized by a
single material parameter (called the hydrogel deformation constant),
although the free-energy function for the hydrogel contains two
material parameters. Comparisons with numerical solutions are also
made and good agreements are found.

\end{abstract}



\section{Introduction}
Gels, known as a cross-linked solution,\cite{hydrogelpolymer1} consist
of a solid three-dimensional network of polymer that spans the volume
of a liquid medium and imbibes the solvent molecules through surface
tension effects. When the solvent happens to be water, the aggregate
is called hydrogel (e.g. edible jelly), which can undergo large and
reversible volumetric deformation by absorbing or expelling water in
response to various external stimuli (e.g. temperature, physical or
chemical stimuli like light and pH). It undergoes a homogeneous
deformation without external mechanical constraint, but an
inhomogeneous and anisotropic one under external constraints (often
present in practice).

This paper deals with a core-shell structure, with a shell of gel
fixed to a hard core of another material (like metal or another
polymer), which defines an inner boundary of the network. Due to the
good properties such as stability, ease of synthesis,
thermalsensitivity and biocompatible nature etc., such a hydrogel
shell has various promising applications including  drug
delivery,\cite{hydrogelkumacheva1,hydrogellee1,hydrogellyon1} medical
devices,\cite{hydrogelguo1} bioseparation\cite{hydrogelkawaguchi1} and
catalysis.\cite{hydrogelbariffs1,hydrogelbariffs2} Some
experiments\cite{hydrogelballauff1,hydrogelballauff2} have been
performed on such a structure in recent years. It was found that there
exists a density fluctuation within the network which indicates the
spatial inhomogeneity. Sometimes, the partial detachment of the shell,
which means the large stress at the inner surface due to the strong
swelling, was observed. Thus a good understanding of equilibrium
swelling states is of crucial importance. However, few analytical
results exist for such inhomogeneous swelling and consequently there
lacks the interpretation of the influence of the material parameters
on the deformation.

Equilibrium theories of heterogeneous substances date back to
Gibbs,\cite{hydrogelgibbs1} who formulated a theory for the
inhomogeneous equilibrium state of large deformation of an elastic
solid in a solvent. Recently extensive studies have concentrated on
the swelling of
gels.\cite{hydrogeldolbow1,hydrogeldolbow2,hydrogeldurning1}
Particularly based on the field theory of Gibbs\cite{hydrogelgibbs1}
and the poroelasticity theory of
Biot,\cite{hydrogelbiot2,hydrogelbiot1} Hong et al.\cite{hydrogelsuo2}
formulated a theory of couple mass transport and deformation in gels
by considering both the mixing and stretching processes, leaving open
the free-energy function. For the specific core-shell structure of
hydrogel, Zhao et al.\cite{hydrogelsuo1} and Hong et
al.\cite{hydrogelsuo3} adopted the free-energy function introduced by
Flory and Rehner\cite{hydrogelflory1} and obtained some numerical
results of the inhomogeneous swelling states, showing large stresses
near the core-shell interface.

The present work is restricted to the equilibrium swelling states
(i.e. the long-time limit) without considering kinetics, of which the
deformation of the network is governed by a boundary-value problem.
The object of this paper is to seek analytical solutions of radially
symmetric deformations for such a core-shell structure, based on the
existing model in Ref.~\refcite{hydrogelsuo1}. Usually, it is very
difficult to obtain analytical solutions for an inhomogeneous state of
a hydrogel due to the nonlinearity caused by the large deformation. In
the case of uniform swellings (the water concentration is uniformly
distributed), a number of analytical solutions have been
obtained.\cite{hydrogelpence1,hydrogelpence2,hydrogelpence3,hydrogeltsai1}
For the present problem, the water concentration is nonuniform, and as
far as we know, analytical solutions for this type of problems are not
available in literature.

Here we intend to construct the asymptotic solutions for the present
problems. Identifying a small parameter in the governing equations, we
analyze the deformations by perturbation methods. For the homogeneous
deformation, by defining a hydrogel deformation constant $\alpha$ we
can express the free stretch in a simple formula. For the general
inhomogeneous deformation, we treat it as a boundary-layer problem of
a nonlinear second-order variable-coefficient differential equation.
It turns out that there is a boundary layer near the hard core, but
the existing method of matched asymptotics does not work for the
present problem. Here we introduce an extended method of matched
asymptotics to construct the analytical solution. More specifically,
this novel methodology involves the introduction of a transition
region besides the usual inner and outer regions and using a series
solution in this region.

This paper is arranged as follows. Section 2 briefly recalls the
formulation of Zhao et al.\cite{hydrogelsuo1} for the hydrogel in the
equilibrium state. We then consider in section 3 the free-swelling
deformation with no external mechanical constraint, and in section 4
we discuss a near free-swelling deformation with the fixed hoop
stretch at the inner surface not far from the free stretch. Section 5
discusses a general inhomogeneous deformation without such a
restriction on the fixed stretch, where an extended method of matched
asymptotics is introduced to construct the analytical solution.
Finally some conclusions are drawn.

\section{Governing Equation}
For the structure of a spherical shell, the spherical symmetric
deformation of the hydrogel is fully specified by a function $r(R)$.
In this section we briefly recall the formulation of Zhao et
al.\cite{hydrogelsuo1} for the hydrogel in the equilibrium state. The
field equation is
\begin{equation}
\label{eq1} \frac{d s_{r}}{d R}+2\frac{s_r-s_\theta}{R}=0,
\end{equation}
where $s_{r},s_{\theta}$ are the nominal stresses in the radial and
circumferential directions respectively.

We adopt the free energy function of the hydrogel first introduced by
Flory and Rehner (see
Ref.~\refcite{hydrogelpolymer2},\refcite{hydrogelflory1}) and follow
the notations in Hong et al.\cite{hydrogelsuo2}
\begin{equation}
\label{eq2_1}
\begin{aligned}
W(\mathbf{F},C)=&W_s(\mathbf{F})+W_m(C) \\
= &\frac{1}{2}NkT[\lambda_1^2+ \lambda_2^2 +\lambda_3^2 -3-2\log
\lambda_1\lambda_2\lambda_3]\\
&- \frac{kT}{v}\left[vC \log
(1+\frac{1}{vC})+\frac{\chi}{1+vC}\right],
\end{aligned}
\end{equation}
where $\mathbf{F}$ is the deformation gradient, $C$ is the nominal
concentration of water (i.e. the number of the water molecules per
reference volume in the current state), $N$ is the number of polymer
chains per reference volume of dry network, $kT$ is the temperature in
the unit of energy ($k$ is the Boltzmann's constant), $\lambda_1,
\lambda_2$ and $\lambda_3$ are the three principal stretches, and $v$
is the volume per solvent molecule (water), $\chi$ is a parameter from
the heat of mixing. The two dimensionless parameters $\chi$ and $vN$
vary in the ranges $0.1-0.5$ and $10^{-2} - 10^{-5}$ respectively
according to  Zhao et al.\cite{hydrogelsuo1} ($1/vN$ actually is the
number of water molecules occupied the same volume of per polymer
chain).

For a spherically symmetrical deformation, it is easy to deduce from
(\ref{eq2_1}) that
\begin{equation}
\label{eq2}
\begin{aligned}
\frac{s_{\theta}}{NkT}&=\lambda_\theta-\lambda_\theta^{-1}
+\frac{\lambda_\theta \lambda_r}{vN}\left[ \log
\frac{vC}{1+vC}+\frac{1}{1+vC}+\frac{\chi}{(1+vC)^2} \right],\\
\frac{s_{r}}{NkT}&=\lambda_r-\lambda_r^{-1}
+\frac{\lambda_\theta^2}{vN}\left[ \log
\frac{vC}{1+vC}+\frac{1}{1+vC}+\frac{\chi}{(1+vC)^2} \right],
\end{aligned}
\end{equation}
where $\lambda_\theta$ and $\lambda_r$ are respectively the stretches
in the circumferential and radial directions, and $vC$ represents the
change in volume of the gel, which are given by
\begin{equation}
\label{eq3} \lambda_\theta=\frac{r}{R},\quad
\lambda_r=\frac{dr}{dR},\quad vC=\lambda_\theta^2 \lambda_r-1.
\end{equation}

Substituting (\ref{eq2}-\ref{eq3}) into (\ref{eq1}), a nonlinear
second-order variable-coefficient differential equation for $r(R)$
arises, which will be solved analytically subjected to suitable
boundary conditions.

In the reference configuration (a water-free and stress-free state),
suppose that the hydrogel shell has the inner and outer radii $A$ and
$B$ respectively. Suppose that in the current configuration (an
equilibrium state immersed in water) the inner surface is attached
with a rigid core and has the radius $r(A)=\lambda_0 A$. At the outer
surface it is supposed that $s_r (B)=0$ or $s_{\theta} (B)=0$.

Since the change in volume $vC$ is relatively large (see Figure 2(a)
in Ref.~\refcite{hydrogelsuo1}), we approximate the term $\log
\frac{vC}{1+vC}$ by the Taylor expansion in terms of $\frac{1}{1+vC}$.
Then, from $(\ref{eq2}-\ref{eq3})$ we have
\begin{equation}
\label{eq5}
\begin{aligned}
\frac{s_{\theta}}{NkT}&=\lambda_\theta-\lambda_\theta^{-1}
-\frac{1-2\chi}{2vN}\frac{1}{\lambda_\theta^3 \lambda_r}-
\frac{1}{3vN}\frac{1}{\lambda_\theta^5 \lambda_r^2}+\cdots,\\
\frac{s_{r}}{NkT}&=\lambda_r-\lambda_r^{-1}
-\frac{1-2\chi}{2vN}\frac{1}{\lambda_\theta^2 \lambda_r^2}-
\frac{1}{3vN}\frac{1}{\lambda_\theta^4 \lambda_r^3}+\cdots.\\
\end{aligned}
\end{equation}
We notice that a small parameter $vN$ appears in the equation, so we
would like to take advantage of this by using perturbation methods to
get approximate analytical solutions for the following three cases.

\section{Explicit Solution for a Free-swelling Deformation}

If the hydrogel swells freely with no external mechanical constraint,
the deformation is homogeneous and isotropic, i.e.
$\lambda_r=\lambda_\theta=\lambda_{free}=(vC_{free}+1)^{1/3}$, which
can be obtained by solving $s_r=0$ (or equivalently $s_\theta=0$).
Now, we shall deduce the explicit asymptotic solution.

Substituting $\lambda:=\lambda_{r}=\lambda_{\theta}$ into
$(\ref{eq5})$ we arrive at
\begin{equation}
\label{eq6} \lambda-\lambda^{-1}
-\frac{1-2\chi}{2vN}\frac{1}{\lambda^4}-
\frac{1}{3vN}\frac{1}{\lambda^7}+\cdots=0.
\end{equation}
Since $vC_{free}$ is large, we also regard $\lambda$ as a large
quantity. From the above equation we can see that the term to balance
the third term, which is large due to the small parameter $vN$, is the
first term $\lambda$. Thus, they should have the same order, which
implies that to the leading order
\begin{equation}
\label{eq7} \lambda = [(1-2\chi)/(2vN)]^{1/5} =: \alpha.
\end{equation}
We call $\alpha$ to be the {\it hydrogel deformation constant}, as we
shall see that this single parameter plays a dominant role for the
deformation. Letting $\lambda=\alpha \tilde{\lambda}$ and seeking a
perturbation expansion solution of $(\ref{eq6})$ in the form
\begin{equation}
\label{eq9}
\tilde{\lambda}=1+\alpha^{-1}\lambda_1+\alpha^{-2}\lambda_2+
O(\alpha^{-3}),
\end{equation}
where $\alpha$ is treated as a large parameter, we obtain the formula
\begin{equation}
\label{eq10} \lambda_{free}=\lambda=\alpha+\frac{1}{5}\alpha^{-1}+
O(\alpha^{-2}).
\end{equation}
We can see that the single parameter $\alpha$, which is a combination
of the original parameters $\chi$ and $vN$, determines the deformation
(up to the order $O(\alpha^{-1})$), i.e., the deformation is not
really two-parameter dependent but rather is mainly one-parameter
dependent.

Thus the current volume per reference volume is
\begin{equation}
1+vC=\lambda^3=\alpha^3+\frac{3}{5} \alpha.
\end{equation}
To the leading order, this result implies that this volume depends on
$1/vN$ by the power $3/5$, which is consistent with a result obtained
before (see eq(13) in Ref.~\refcite{hydrogelflory1}). Here, the
correction term ($\frac{3}{5} \alpha$) is also provided.

Actually $\lambda_{free}$ can be calculated numerically directly from
the formula in $(\ref{eq2})$. For several sets of parameters we
compare the $\lambda_{free}$ values according to our explicit solution
and the numerical solution in the following table:
\begin{table}[ht]
\tbl{Comparison of explicit solution and the numerical solution for
$\lambda_{free}$.} {\begin{tabular}{@{}ccccc@{}} \toprule
$(vN,\chi)$& $\alpha$ & numerical solution & explicit solution & error\\
\hline $(10^{-2},0.2)$ &1.97435 & 2.12537 &2.07565 & 2.3\%\\
\hline $(10^{-3},0.2)$ & 3.12913 &3.21502 &3.19305 &0.68\%\\
\hline $(10^{-4},0.2)$ & 4.95934 &5.00872 &4.99967 &0.18\%\\
\hline $(10^{-5},0.2)$ & 7.86003 &7.88911 &7.88548 & 0.05\%\\
\hline $(\frac{2}{3}\times10^{-4},0.3)$ & 4.95934
&5.01302 &4.99967 &0.27\% \\
\botrule
\end{tabular}}
\end{table}

We can see that the very simple formula $(\ref{eq10})$ for
$\lambda_{free}$ agrees with the numerical solution very well. As $vN$
or $\chi$ decreases, $\alpha$ increases, and thus the explicit
solution becomes more accurate. However, even when $\alpha^{-1}$ is
not so small the explicit solution gives a very good result already
(say, in the case of row one $\alpha^{-1}=0.5065$ and the error is
only 2.3\%). This often happens for a perturbation expansion solution:
In theory one needs that the small parameter tends to zero but in
practice the result can be valid even when the parameter is not so
small.

The fifth row should be compared with the third row. Although the
values for $vN$ and $\chi$ are different, the single parameter
$\alpha$ has the same value in the two cases. It can be seen that the
values of $\lambda_{free}$ according to the numerical solution are
also almost the same.

\section{Analytical Solution for a Near Free-swelling Deformation}

In practice mechanical constraints at the outer and inner surfaces may
be present and as a result the deformation is inhomogeneous. In this
section we consider the case that the inner surface $R=A$ has a fixed
radial displacement $r(A)-A=\lambda_0 \cdot A - A$ (i.e., the stretch
$\lambda_\theta=\lambda_0$) and the outer surface is still stress-free
in the radial direction. It is further supposed that
$|\lambda_0-\lambda_{free}|\ll\alpha$ or $|\lambda_0-\alpha|\ll\alpha$
for a large $\alpha$. For this problem, one would expect that the
deformation, although inhomogeneous, is close to a free-swelling one
as $\lambda_0$ is close to $\lambda_{free}$. Now, we proceed to
construct the explicit analytical solution.

For a deformation close to that of a free swelling state, to the
leading order, the deformation should be given by $r(R)=\alpha R$. We
make the following transformation:
\begin{equation}
\label{eq11} r(s)=\alpha u(s) R, \quad s=\frac{R-A}{B-A},
\end{equation}
where $s$ is used as the independent variable of $u$ and $r$ in order
to convert the domain $[A,B]$ to the unit interval $[0,1]$. Then, from
(\ref{eq5}) and $(\ref{eq1})$ we arrive at
\begin{equation}
\label{eq13}
\begin{aligned}
&\left[1+\frac{2}{u^2(s) [u(s)+(s+a)u'(s)]^3}\right](s+a)u''(s)+\\
&\left[4+\frac{4}{u^3(s) [u(s)+(s+a)u'(s)]^2}+\frac{4}{u^2(s)
[u(s)+(s+a)u'(s)]^3}\right]u'(s)+\\
&\alpha^{-2}\left[\frac{2u'(s)+(s+a)u''(s)}{[u(s)+(s+a)u'(s)]^2}
+\frac{2u'(s)}{u(s)[u(s)+(s+a)u'(s)]}\right]+O(\alpha^{-3})=0,
\end{aligned}
\end{equation}
where $a=A/(B-A)$ is the ratio of the inner radius to the shell
thickness.

At the outer surface $s=1$, the outer boundary condition $s_r(1)=0$
implies that
\begin{equation}
\label{eq14}
\begin{aligned}
&u(1)+(a+1) u'(1)-\frac{1}{u^{2}(1) \left[u(1)+(a+1)
u'(1)\right]^2}\\
&-\alpha^{-2}\frac{1}{u(1)+(a+1) u'(1)}+O(\alpha^{-3})=0.
\end{aligned}
\end{equation}
At the inner surface $s=0$, the boundary condition becomes
\begin{equation}
\label{eq15} r(0)=\lambda_0 A \quad \Rightarrow
u(0)=\frac{\lambda_0}{\alpha} :=1+\alpha^{-1}\lambda_0^\ast,
\end{equation}
where $\lambda_0^\ast = \lambda_0-\alpha$ is regarded as an $O(1)$
quantity (so that $|\lambda_0-\alpha|\ll\alpha$).

Next we seek a regular perturbation expansion solution by considering
the parameter $\alpha$ to be large. Since to the leading order $u(s)$
should be $1$ for a near free-swelling deformation, we let
\begin{equation}
\label{eq16} u(s)=1+\alpha^{-1}u_1(s)+\alpha^{-2}u_2(s)+\cdots.
\end{equation}
At $O(1)$, equation $(\ref{eq13})$ and boundary conditions
$(\ref{eq14}-\ref{eq15})$ are automatically satisfied. At
$O(\alpha^{-1})$, we have from equation $(\ref{eq13})$ that
\begin{equation}
\label{eq17} (s+a)u_1''(s)+4u_1'(s)=0.
\end{equation}
Solving this equation and further using boundary conditions
$(\ref{eq14}-\ref{eq15})$, we obtain
\begin{equation}
\label{eq18} u_1(s)=c_1(s+a)^{-3}+c_2,
\end{equation}
where
\begin{equation}
\label{eq19}
\begin{aligned}
c_1=\frac{5a^3(1+a)^3\lambda_0^\ast}{5(1+a)^3+4a^3},\quad
c_2=\frac{4a^3\lambda_0^\ast}{5(1+a)^3+4a^3}.
\end{aligned}
\end{equation}
At $O(\alpha^{-2})$, from equation $(\ref{eq13})$ we obtain
\begin{equation}
\label{eq20} (s+a)u_2''(s)+4u_2'(s)=-12c_1^2(s+a)^{-7}.
\end{equation}
Solving this equation and further using boundary conditions
$(\ref{eq14}-\ref{eq15})$, we obtain
\begin{equation}
\label{eq21} u_2(s)=d_1(s+a)^{-3}+d_2-\frac{2}{3}c_1^2(s+a)^{-6},
\end{equation}
where
\begin{equation}
\label{eq22} d_1=\frac{a^3(1+a)^3(5M_1-M_2)}{5(1+a)^3+4a^3},\quad
d_2=\frac{4a^3M_1+(1+a)^3M_2}{5(1+a)^3+4a^3},
\end{equation}
and
\begin{equation}
\label{eq23} M_1=\frac{2}{3}c_1^2a^{-6},\quad
M_2=1-\frac{5c_1^2}{3(1+a)^6}- \frac{10c_1c_2}{(1+a)^3}+10c_2^2.
\end{equation}

By transferring back to the original variable $R$, up to
$O(\alpha^{-1})$, the solution is given by
\begin{equation}
\label{eq24} \tilde{r}(R)=\alpha \tilde{R}+\frac{c_1}{a^3
\tilde{R}^2}+c_2 \tilde{R}+\alpha^{-1} \left(\frac{d_1}{a^3
\tilde{R}^2}+d_2 \tilde{R} -\frac{2c_1^2}{3a^6 \tilde{R}^5}\right),
\end{equation}
where $\tilde{r}=r/A$ and $\tilde{R}=R/A$. We point out that $c_1,
c_2, d_1$ and $d_2$ only depend on the geometric parameter $a$ and
$\lambda_0^\ast$.

The analytical solution can provide a lot insight information. First,
once again we can see that the deformation is mainly characterized by
the single hydrogel deformation constant $\alpha$. Next, we shall
present the analytical formulas for the physical quantities at the
inner and outer surfaces. At the inner surface $R=A$, from the
analytical solution, the following simple formulas (valid up to
$O(1)$) can be immediately induced:
\begin{equation}
\label{eq25}
\begin{array}{ll}
\displaystyle \lambda_\theta=\alpha+\lambda_0^\ast ,\quad
&\displaystyle \lambda_r=\alpha-\frac{2[5(1+a)^3-2a^3]
\lambda_0^\ast}
{5(1+a)^3+4a^3},\\
\displaystyle \frac{s_\theta}{NkT}=\frac{10[(1+a)^3+2a^3]
\lambda_0^\ast} {5(1+a)^3+4a^3}, \quad &\displaystyle
\frac{s_r}{NkT}=-\frac{20[(1+a)^3-a^3]  \lambda_0^\ast}
{5(1+a)^3+4a^3}.
\end{array}
\end{equation}
At the outer surface $R=B$, we have
\begin{equation}
\label{eq26}
\begin{array}{ll}
\displaystyle \lambda_\theta=\alpha+\frac{9 a^3 \lambda_0^\ast}
{5(1+a)^3+4a^3}, \quad &\displaystyle \lambda_r=\alpha-\frac{6 a^3
\lambda_0^\ast}
{5(1+a)^3+4a^3},\\
\displaystyle \frac{s_\theta}{NkT}=\frac{30 a^3 \lambda_0^\ast}
{5(1+a)^3+4a^3}, \quad &\displaystyle \frac{s_r}{NkT}=0.
\end{array}
\end{equation}
The stress values at the inner surface are of particular interest as
debonding may happen there. We notice that at the inner surface both
stress values are proportional to the value
$\lambda_0^\ast=\lambda_0-\alpha$, the difference between the given
stretch and the hydrogel deformation constant $\alpha$, with the
proportional constants dependent on the single geometric parameter
$a$, the ratio of the inner radius to the shell thickness. At the
outer surface, $s_\theta$ is not zero, rather it is an $O(1)$ quantity
proportional to $\lambda_0^\ast$. This implies that certain stress in
the circumferential direction has to be applied to maintain this
spherically symmetric deformation.

To further examine the influence of the geometric parameter $a$, we
consider two special situations: $a\ll1$ and $a\gg1$, which correspond
to the cases of the shell being very thick and very thin (relative to
the inner radius) respectively.

For $a\ll 1$,  at $R=A$ we have
\begin{equation}
\label{eq27} \lambda_\theta=\alpha+\lambda_0^\ast=\lambda_0 ,\quad
\lambda_r\approx\alpha-2 \lambda_0^\ast,\ \
\frac{s_\theta}{NkT}\approx2 \lambda_0^\ast, \quad
\frac{s_r}{NkT}\approx-4  \lambda_0^\ast,
\end{equation}
and at $R=B$ we have
\begin{equation}
\label{eq28} \lambda_\theta\approx\alpha+\frac{9}{5} a^3
\lambda_0^\ast,\ \lambda_r\approx\alpha-\frac{6}{5} a^3
\lambda_0^\ast,\
 \frac{s_\theta}{NkT}\approx6 a^3 \lambda_0^\ast,\  \frac{s_r}{NkT}=0.
\end{equation}
In this case, we see that at the inner surface the magnitude of the
stress $s_r$ is twice that of $s_\theta$ and their signs are opposite.
Also, $s_\theta$ is very small at the outer surface (as $a\ll 1$),
which implies that little stress in the circumferential direction
needs to be applied.

For $a\gg 1$, at $R=A$ we have
\begin{equation}
\label{eq29} \lambda_\theta=\alpha+\lambda_0^\ast=\lambda_0 , \
\lambda_r\approx\alpha-\frac{2}{3} \lambda_0^\ast,\
\frac{s_\theta}{NkT}\approx \frac{10}{3} \lambda_0^\ast, \
\frac{s_r}{NkT}\approx-\frac{20}{3a} \lambda_0^\ast\approx0,
\end{equation}
and at $R=B$ we have
\begin{equation}
\label{eq30} \lambda_\theta\approx\alpha+\lambda_0^\ast=\lambda_0 , ~~
\lambda_r\approx\alpha-\frac{2}{3} \lambda_0^\ast,~~
\frac{s_\theta}{NkT}\approx \frac{10}{3} \lambda_0^\ast,~~
\frac{s_r}{NkT}=0.
\end{equation}
In this case, the stresses and stretches at the inner and outer
surfaces are approximately same, which are somehow expected for a thin
shell. In contrast to the first case, the stretches $\lambda_\theta$
and $\lambda_r$ at the outer surface differ from $\alpha$ (or
$\lambda_{free}$) by an $O(1)$ quantity, and the stress $s_\theta$ at
the outer surface is not small but an $O(1)$ quantity, which means
that an $O(1)$ stress needs to be applied at the outer surface for
such a deformation.

The nonlinear second-order variable-coefficient differential equation
$(\ref{eq1})$ with the boundary conditions $r(A)=\lambda_0 A$ and $s_r
(B)=0$ can be solved by using a numerical method. To examine the
validity of our analytical solution obtained above, we use a shooting
method to get the numerical solution and then compare it with the
analytical one. In Figure 1, the solution curves according to the two
methods are plotted.
\begin{figure}[H]
\begin{center}
\subfigure[ The parameter values are $vN=10^{-4}$ and $\chi=0.2$
($\alpha=4.95934$), $B=3A$ and $\lambda_0=4$ (correspondingly
$\lambda_0^\ast\approx
-1$).]{\includegraphics[width=2.4in]{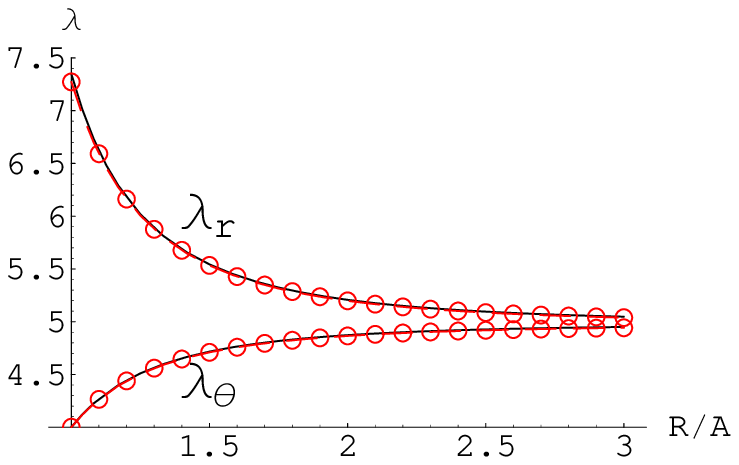}} \hspace{0.05in}
\subfigure[The parameter values are $vN=10^{-4}$ and $\chi=0.2$
($\alpha=4.95934$), $B=3A$ and $\lambda_0=3.5$ (correspondingly
$\lambda_0^\ast\approx
-1.5$).]{\includegraphics[width=2.4in]{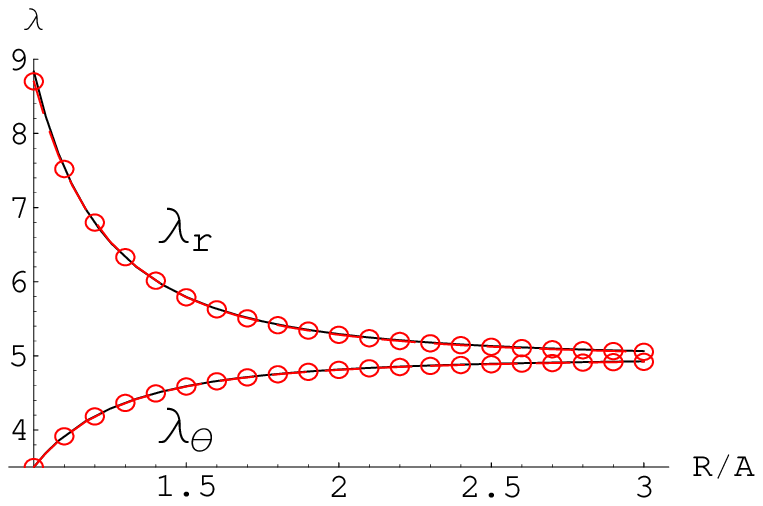}} \subfigure[ The
parameter values are $vN=10^{-4}\times2/3$ and $\chi=0.3$
($\alpha=4.95934$), $B=3A$ and $\lambda_0=4$ (correspondingly
$\lambda_0^\ast\approx
-1$).]{\includegraphics[width=2.4in]{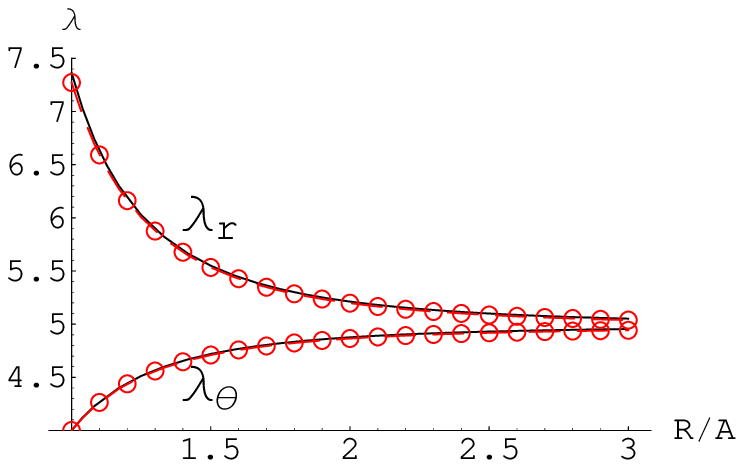}} \hspace{0.05in}
\subfigure[The parameter values are $vN=10^{-4}\times2/3$ and
$\chi=0.3$ ($\alpha=4.95934$), $B=3A$ and $\lambda_0=3.5$
(correspondingly $\lambda_0^\ast\approx -1.5$).]{
\includegraphics[width=2.4in]{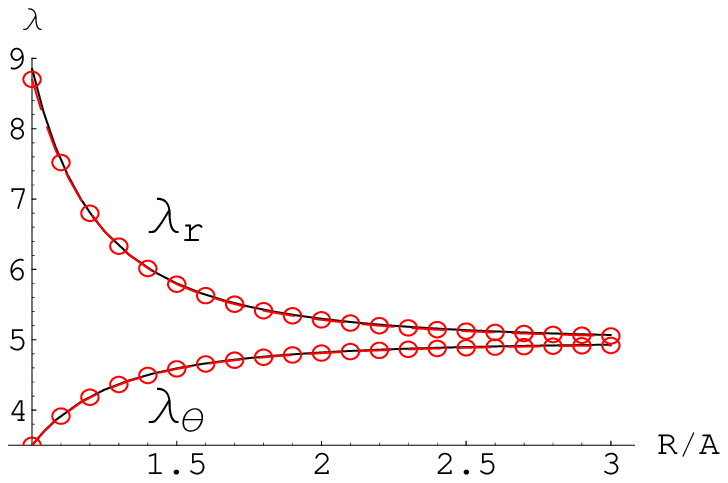}}
\end{center}
\caption{Distributions of the stretches (lines -o- are the analytical
solutions and black solid lines are the numerical solutions).}
\end{figure}

For the chosen geometric parameter in Figure 1, we have $a=0.5$.
Although the value of $\alpha$ is not very large, it can be seen that
the analytical solution agrees with the numerical one very well.
Actually, the maximum relative errors\footnote{it is defined as the
maximum error divided by the maximum value i.e. $\max
|\hat{y}-y|/\max|y|$} of $(\lambda_\theta,\lambda_r)$ are only about
$(0.2\%,1\%)$,$(0.2\%,1.5\%)$,$(0.3\%,1.3\%)$ and $(0.2\%,1.7\%)$
respectively for the four figures.

We also point out that $(vN, \chi)$ have different values in Figures
$(a,b)$ and Figures $(c,d)$, but the $\alpha$ value is the same in all
cases. So, the analytical solutions in Figures $(a,c)$ and Figures
$(b,d)$ are the same respectively. The agreement between the
analytical solutions and numerical ones show that the deformation is
mainly determined by the single hydrogel deformation constant
$\alpha$, although the free-energy function contains two material
constants $(vN,\chi)$.

Normally when the outer boundary condition $s_r(B)=0$ is used, the
stress $s_\theta$ is not 0 but an $O(1)$ quantity (see
$(\ref{eq26})_{3}$). Now, we consider the case that $s_\theta(B)=0$
instead of $s_r(B)=0$. In this case, the boundary condition
$(\ref{eq14})$ is replaced by
\begin{equation}
\label{eq31} u(1)-\frac{1}{u^3(1) \left[u(1)+(a+1)
u'(1)\right]}-\alpha^{-2} \frac{1}{u(1)} +O(\alpha^{-3})=0.
\end{equation}
One can proceed to construct the perturbation expansion solution as
before. The solution expression is still given by $(\ref{eq24})$ but
now the expressions for the constants are replaced by
\begin{equation}
\label{eq32} c_1=\frac{5a^3(1+a)^3\lambda_0^\ast}{5(1+a)^3-2a^3},\quad
c_2=\frac{-2a^3\lambda_0^\ast}{5(1+a)^3-2a^3},
\end{equation}
\begin{equation}
\label{eq33}
\begin{aligned}
d_1=\frac{a^3(1+a)^3(5M_1-M_2)}{5(1+a)^3-2a^3},\quad
d_2=\frac{(1+a)^3M_2-2a^3M_1}{5(1+a)^3-2a^3},
\end{aligned}
\end{equation}
where $M_1$ is given by $(\ref{eq23})_1$ and
\begin{equation}
\label{eq34} M_2=1+\frac{10c_1^2}{3(1+a)^6}+
\frac{5c_1c_2}{(1+a)^3}+10c_2^2.
\end{equation}
Simple analytical formulas can also be obtained for the stretches and
stresses at the inner and outer surfaces, and here we omit the
details.

\section{Analytical Solution for a General Inhomogeneous Deformation}
In the previous section, we have assumed that at the inner surface the
given stretch $\lambda_0$ satisfies the constraint
$|\lambda_0-\alpha|\ll \alpha$. In that case, basically the governing
equation can be linearized around $r=\alpha R$ so the analytical
solution can be obtained by solving linear differential equations.
Now, we shall proceed to construct the solution without the above
constraint. Instead, it is supposed that the stretch $\lambda_0$ at
$R=A$ is far away from $\alpha$ such that $\lambda_0$ is an $O(1)$
quantity. For this problem, one cannot avoid to deal with some
nonlinear second-order variable-coefficient differential equation(s).

As mentioned before, for a hydrogel the material constant $vN$ is
small, so we always take $\alpha$ as a large parameter or
$\alpha^{-1}$ as a small parameter. In general, one cannot solve a
nonlinear differential equation analytically. However, if a small
parameter is present in the equation, sometimes one can use singular
perturbation methods to construct asymptotic solutions. But, for those
methods to work, usually the equation should become degenerate as the
small parameter tends to zero, say, it becomes a linear equation or it
becomes a first-order equation instead of the original second-order
equation. For the present problem governed by $(\ref{eq13})$, we see
that as $\alpha^{-1}$ tends to zero the leading-order equation is
still a complicated nonlinear second-order variable-coefficient
equation. This shows that the existing singular perturbation methods
do not work for this equation. Here, we introduce a novel methodology,
which is an extension of the method of matched asymptotics, to
construct the analytical solution.

We consider the case that the thickness of the shell is relatively
large (an explicit restriction will be provided later on). We first
make some observations on the solution structure. For a thick shell,
the boundary condition at the inner surface should not influence a
region some distance away from it (we assume that the St. Venant's
principle applies). So, there is a region containing the outer
boundary point in which the deformation is near a free-swelling one as
the stress-free condition can then be satisfied automatically (to the
leading order). We call this region to be the outer region. When
$|\lambda_0-\alpha|\ll \alpha$, we see from Figure 1 that in a region
near the inner surface the stretches change rapidly. It is reasonable
to expect that when $\lambda_0=O(1)$ the stretches also change rapidly
in this region. In other words, there is a boundary layer region near
the inner surface, and we call this region to be the inner region.
This kind of structure can also be seen from the numerical solutions
obtained in Zhao et al..\cite{hydrogelsuo1} In the standard technique
of matched asymptotics\cite{perturbation1} only an inner region and an
outer region exist and the equation in the latter region is one-order
less than that in the former region. By solving the equations in both
regions separately and then matching the two solutions together to
determine the integration constants, the asymptotic solution can be
obtained. However, for the present problem, the leading-order equation
in the outer region, which can be obtained by setting $\alpha^{-1}=0$
in $(\ref{eq13})$, is still a second-order differential equation, so
one cannot simply match the solutions in the inner region and outer
region directly. To connect them there should be a third region in
between, which will be called a transition region. We shall use this
transition region to connect the outer and inner regions. However, a
major difficulty arises: In this region one has to deal with the full
nonlinear second-order variable-coefficient differential equation,
which is not solvable analytically! We shall overcome this difficulty
by using a series expansion for the solution in this region whose
interval should be small. The details are described below.
\vspace{0.5cm}

{\it (a) Solution in the outer region}

First we consider the outer solution. The governing equation is still
$(\ref{eq13})$, and the boundary condition $(\ref{eq14})$ can still be
used in the outer region. As mentioned before, it is expected that in
this region the deformation is near a free-swelling one. Therefore, we
seek a perturbation expansion solution of the form
\begin{equation}
\label{eq40} u(s)=1+\alpha^{-2} u_1(s)+\cdots.
\end{equation}
Here, the second-order term is set to be $O(\alpha^{-2})$, to be
consistent with the governing equation $(\ref{eq13})$ and boundary
condition $(\ref{eq14})$. We substitute this expansion into
$(\ref{eq13})$ and $(\ref{eq14})$. At $O(1)$, they are automatically
satisfied. At $O(\alpha^{-2})$, we find that $u_1(s)$ satisfies
$(\ref{eq17})$, and the solution expression is
\begin{equation}
\label{eq41} u_1(s)=C_3(s+a)^{-3}+C_2,
\end{equation}
where $C_2$ and $C_3$ are two integration constants. By further using
$(\ref{eq14})$, we obtain
\begin{equation}
\label{eq42} u_1(s)=\frac{(5C_2-1)(1+a)^3}{4(s+a)^3}+C_2.
\end{equation}
To sum up, the outer solution is given by
\begin{equation}
\label{eq43} \tilde{r}_{out}(s):=\frac{r_{out}(s)}{A}=\alpha
\frac{s+a}{a}+ \alpha^{-1} \left[\frac{(5C_2-1)(1+a)^3}{4a(s+a)^2} +
\frac{C_2(s+a)}{a} \right],
\end{equation}
where $C_2$ is to be determined. \vspace{0.5cm}

{\it (b) Solution in the inner region}

Next we consider the inner region (i.e. boundary layer), we should
examine the full equation $(\ref{eq13})$. To simplify the equation we
introduce the variable $\bar{r}$ by
\begin{equation}
\label{eq44} \frac{r}{A}=\tilde{r}=\alpha \bar{r}(s)\quad \Rightarrow
\quad u(s)=\frac{a\bar{r}(s)}{s+a}.
\end{equation}
Then the equation (\ref{eq13}) becomes
\begin{equation}
\label{eq45}
\begin{aligned}
&\left[a^2+\frac{2(s+a)^2}{a^3\bar{r}^{2}(s) [\bar{r}'(s)]^3}\right]
\bar{r}''(s) +\frac{2a^2\bar{r}
'(s)}{(s+a)}-\frac{2a^2\bar{r}(s)}{(s+a)^2}
-\frac{4(s+a)}{a^3\bar{r}^2(s) [\bar{r} '(s)]^2}\\
&+\frac{4(s+a)^2}{a^3\bar{r}^3(s) \bar{r}
'(s)}+\alpha^{-2}\left[\frac{(a+s) \bar{r}''(s)-2 \bar{r}'(s)}{(a+s)
[\bar{r}'(s)]^2}+\frac{2}{\bar{r}(s)}\right]=0.
\end{aligned}
\end{equation}

Suppose that in this region the maximum value
$\bar{r}_{max}=O(\alpha^{-k})$ ($k$ is to be determined) and we write
$ \bar{r}=\alpha^{-k} \hat{r}$. We note that the value of $\bar{r}$ at
the inner surface is $O(\alpha^{-1})$ due to the condition
$\tilde{r}=\lambda_0=O(1)$. $\bar{r}_{max}$ should be much larger than
this value due to the rapid increase of $\bar{r}$ in the boundary
layer region. Thus a restriction is $k<1$. To reflect the rapid change
of $\bar{r}$ in the boundary layer, we introduce the stretching
coordinate $X=s/(a \epsilon)$, where $a \epsilon$, the parameter
characterizing the thickness of the boundary layer, is to be
determined. Making this change of variables to equation
$(\ref{eq45})$, according to the Van Dyke's principle of least
degeneracy,\cite{perturbation2} we find $\epsilon=\alpha^{-5k/3}$.
Since $\epsilon$ should be small, we need $k>0$. And, the equation
becomes

\begin{equation}
\label{eq47} \left[1+\frac{2}{\hat{r}^2 \hat{r}_{_X}^3}\right]
\hat{r}_{_{XX}}+\frac{4}{\hat{r}^3
\hat{r}_{_X}}+O(\epsilon)+O(\epsilon^2)+O(\epsilon^2 \alpha^{2k-2})=0,
\end{equation}
where we denote $\hat{r}_{_X}=\frac{d\hat{r}}{dX}$ to distinguish from
$\hat{r}'(s)$. Since $0<k<1$, $O(\alpha^{2k-2})$ is small. If one uses
the Van Dyke's principle of least degeneracy for the $O(\epsilon^3)$
equation, it is required that $O(\epsilon^2
\alpha^{2k-2})=O(\epsilon^3)$, i.e., $k=6/11$. Then, the boundary
layer thickness parameter $a\epsilon=a\alpha^{-10/11}$, which needs to
be small, say, $a\epsilon<0.15$. Thus, a restriction is $a<0.15
\alpha^{10/11}$.

Multiplying both sides by $\hat{r}_{_X}$ and integrating once, we
obtain (to the leading order)
\begin{equation}
\label{eq48} \hat{r}_{_X}^2-\frac{4}{\hat{r}^2 \hat{r}_{_X}}=C_1,
\end{equation}
where $C_1$ is the integration constant. This is a first-order
differential equation. With the boundary condition
$\tilde{r}(0)=\lambda_0$, theoretically there is only one constant
$C_1$ to be determined. Actually, the solution $\hat{r}_{in}$ of the
above equation can be represented by
\begin{equation}
\label{eq49} X= \int_{\hat\lambda_0}^{\hat{r}_{in}}
\frac{dy}{f(y;C_1)},
\end{equation}
where $\hat\lambda_0:=\lambda_0 \alpha^{-5/11}$ is a known constant
and $f$ is the root of the cubic algebraic equation
\begin{equation}
\label{eq50} F(f)=f^3 -C_1 f - \frac{4}{y^2} = 0.
\end{equation}
For the present problem we require $f=\hat{r}_{_X}>0$ with
$y=\hat{r}>0$. It is easy to show that equation $(\ref{eq50})$ has one
and only one positive root, which is given by
\begin{equation}
\label{eq51} f(y;C_1)=\begin{cases} -\frac{C_1 y^{2/3}}{3 \sqrt[3]{2}
\left[\sqrt{1-C_1^3y^4/108}-1\right]^{1/3}}-\frac{\sqrt[3]{2}
\left[\sqrt{1-C_1^3y^4/108}-1\right]^{1/3}}{y^{2/3}}, & C_1<0,\\
\sqrt[3]{4}/y^{2/3}, & C_1=0,\\
\frac{C_1 y^{2/3}}{3
\sqrt[3]{2}\left[1-\sqrt{1-C_1^3y^4/108}\right]^{1/3}}+\frac{\sqrt[3]{2}
\left[1-\sqrt{1-C_1^3y^4/108}\right]^{1/3}}{y^{2/3}}, & C_1>0.\\
\end{cases}
\end{equation}

In summary, the inner solution is provided by $(\ref{eq49})$ and
$(\ref{eq51})$ with one constant $C_1$ to be determined.
\vspace{0.5cm}

{\it (c)Solution in the transition region}

Now, we consider the transition region, which is used to connect both
the inner and outer regions. Since $\tilde{r}$ (or $\lambda_{\theta}$)
is $O(\alpha^{5/11})$ and $O(\alpha)$ respectively for the inner and
outer regions, such a transition region is needed to get the whole
solution in the whole interval. The independent variable $s$ is in the
interval $[0,1]$, and we schematically represent the three regions in
Figure 2. In this figure, the transition region is represented by
$[s_0-\Delta, s_0+\Delta]$, where $s_0$ and $\Delta$ are to be
determined.
\begin{figure}[H]
\begin{center}
\setlength{\unitlength}{1mm}
\begin{picture}(100,25)
\thicklines \put(10,10){\line(1,0){80}}
\multiput(20,7)(60,0){2}{\line(0,1){6}}
\multiput(20,7)(0,6){2}{\line(1,0){2}}
\multiput(80,7)(0,6){2}{\line(-1,0){2}}
\multiput(33,9)(13,0){2}{\line(0,1){2}} \put(20,3){$0$}
\put(80,3){$1$} \put(26,3){\small{$s_0-\Delta$}}
\put(43,3){\small{$s_0+\Delta$}} \put(10,15){inner region}
\put(60,15){outer region} \put(30,20){transition region}
\put(40,17){\vector(0,-1){4}}
\end{picture}
\end{center}
\caption{The geometric representation of the three regions.}
\end{figure}
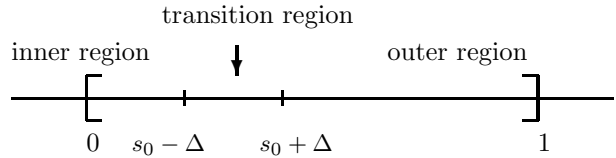

In this region, we should use the full equation $(\ref{eq45})$, and we
have (to the leading order)
\begin{equation}
\label{eq52}
\begin{aligned}
&\left[a^2+\frac{2(s+a)^2}{a^3\bar{r}^{2}(s) [\bar{r}'(s)]^3}\right]
\bar{r}''(s)
+\frac{2a^2\bar{r} '(s)}{(s+a)}-\frac{2a^2\bar{r}(s)}{(s+a)^2}\\
&-\frac{4(s+a)}{a^3\bar{r}^2(s) [\bar{r} '(s)]^2}+
\frac{4(s+a)^2}{a^3\bar{r}^3(s) \bar{r} '(s)}=0.
\end{aligned}
\end{equation}
This is a nonlinear second-order variable-coefficient differential
equation, which appears to be not solvable analytically! To proceed
further, we observe the following: The whole interval for $s$ is
$[0,1]$, which is divided into three regions: outer region, transition
region and inner region. Since usually the outer region is large in a
singular perturbation problem (this is also evident from the numerical
solutions in Zhao et al.\cite{hydrogelsuo1}), the transition region
should only occupy a small subinterval of $[0,1]$. Thus, for $s$ in
the small subinterval $[s_0-\Delta, s_0+\Delta]$ the solution of the
above nonlinear equation can be expanded as a series (as long as $\bar
r (s)$ is sufficiently smooth):
\begin{equation}
\label{eq53}
\begin{aligned}
\bar{r}_{tran}(s)=r_0+r_1 (s-s_0)+r_2 (s-s_0)^2+r_3 (s-s_0)^3+\cdots,
\end{aligned}
\end{equation}
where $r_i (i=0,1,2,3)$ together with $s_0$ need to be determined.
Substituting this expansion into equation $(\ref{eq52})$, the left
hand side becomes a series of $(s-s_0)$. All the coefficients of
$(s-s_0)^n (n=0,1,2,3,\cdots)$ should be zero. From the coefficients
of $(s-s_0)^0$ and $(s-s_0)$, we can obtain two algebraic relations
among the undetermined coefficients, which are represented as
\begin{equation}
\label{eq54}
\begin{aligned}
&f_1(s_0,r_0,r_1,r_2)=0, \\
&f_2(s_0,r_0,r_1,r_2,r_3)=0,
\end{aligned}
\end{equation}
where the lengthy expressions of $f_1$ and $f_2$ are omitted. To have
enough relations for the determination of all constants, we need to
relate the transition solution to the outer and inner
solutions.\vspace{0.5cm}

{\it (d) Determination of the constants through connection
conditions}

We have obtained the solution expressions in the outer region, inner
region and transition region (see equations
$(\ref{eq43}),(\ref{eq49})$ and $(\ref{eq53})$). Each of the outer and
inner solutions contains one constant and the transition solution
contains five constants. The subinterval $[s_0-\Delta, s_0+\Delta]$
also needs to be found, so we have another constant $\Delta$ to
determine. Besides equations $(\ref{eq54})_{1,2}$, we need another six
relations for the eight constants $C_1, C_2, s_0, \Delta,
r_i(i=0,1,2,3)$, which can be obtained by requiring $r,r',r''$ are all
continuous at $s_0-\Delta$ and $s_0+\Delta$, i.e.
\begin{equation}
\label{eq55}
\begin{aligned}
&r_{in}(s)=r_{tran}(s),\ r_{in}'(s)=r_{tran}'(s),\
r_{in}''(s)=r_{tran}''(s),\quad \  {\rm at} \ s=s_0-\Delta,\\
&r_{tran}(s)=r_{out}(s),\ r_{tran}'(s)=r_{out}'(s),\
r_{tran}''(s)=r_{out}''(s),\ {\rm at} \ s=s_0+\Delta.
\end{aligned}
\end{equation}
To reduce the above six relations into two relations, by using the
solution expression $(\ref{eq53})$ we rewrite them as

\begin{equation}
\label{eq56}
\begin{aligned}
&r_0-r_1 \Delta+r_2 \Delta^2- r_3 \Delta^3
+O(\Delta^4)=\tilde{r}_{in}(s_0-\Delta; C_1)/\alpha,  \\
&r_1-2 r_2 \Delta+ 3 r_3 \Delta^2
+O(\Delta^3)=\tilde{r}'_{in}(s_0-\Delta; C_1)/\alpha, \\
&2 r_2 -6 r_3 \Delta +O(\Delta^2)=\tilde{r}''_{in}(s_0-\Delta; C_1)/\alpha, \\
&r_0+r_1 \Delta+r_2 \Delta^2+ r_3 \Delta^3
+O(\Delta^4)=\tilde{r}_{out}(s_0+\Delta; C_2)/\alpha, \\
&r_1+2 r_2 \Delta+ 3 r_3 \Delta^2
+O(\Delta^3)=\tilde{r}'_{out}(s_0+\Delta; C_2)/\alpha,  \\
&2 r_2 +6 r_3 \Delta +O(\Delta^2)=\tilde{r}''_{out}(s_0+\Delta;
C_2)/\alpha.
\end{aligned}
\end{equation}
By some simple manipulations, $r_{j} (j=0,1,2,3)$ can be eliminated,
and as a result two equations for the four constants $s_0, \Delta,
C_1, C_2$ are obtained
\begin{equation}
\label{eq59}
\begin{aligned}
&-3 \tilde{r}_{in}+3 \tilde{r}_{out}+2 \Delta \left(-3
\tilde{r}_{out}'+\Delta \tilde{r}_{in}''+2 \Delta
\tilde{r}_{out}''\right)=0,\\
&-3 \tilde{r}_{in}+3 \tilde{r}_{out}-2 \Delta \left(3
\tilde{r}_{in}'+2 \Delta \tilde{r}_{in}''+\Delta
\tilde{r}_{out}''\right)=0.
\end{aligned}
\end{equation}
where the subscripts ``in" and ``out" represent the value at
$s_0-\Delta$ and $s_0-\Delta$ respectively. Another two equations for
$s_0, \Delta, C_1, C_2$ are provided by $(\ref{eq54})_{1,2}$. By the
Newton's method, these constants can be easily found.

To get the solution curve, we take the parameters $vN=10^{-4}$ and
$\chi=0.2 $, which yields that $\alpha \approx 5$. For the geometrical
parameter we choose two different values $B=3A$ and $4A$ (i.e.,
$a=0.5$ and $1/3$). The stretch at the inner surface is chosen to be
$\lambda_0=1.077$ (which is an $O(1)$ quantity). For such parameters,
by solving the system of 4 algebraic equations mentioned above, we
find
\begin{equation}
\label{eq60}
\begin{aligned}
&a=0.5: \ s_0=0.2127, \Delta=0.06265, C_1=-0.008259,C_2=-0.2449, \\
&a=1/3: \ s_0=0.1670, \Delta=0.04386, C_1=0.0008190,C_2=0.01042.
\end{aligned}
\end{equation}
For such parameters, we have $\epsilon=\alpha^{-10/11}=0.2332$. The
parameter $a\epsilon$, a measure of the magnitude of the boundary
layer thickness has the values about $0.12$ and $0.08$ for $a=0.5$ and
$1/3$ respectively, which are consistent with the values of $
s_0-\Delta$ (0.15 and 0.12). The subintervals of the transition region
$[0.150,0.275]$ and $[0.123,0.211]$ (for $a=0.5$ and $a=1/3$
respectively) are indeed small, as observed before. For such small
intervals, the series solution $(\ref{eq53})$ should be very accurate.

Finally we compare our analytical solution with the numerical one
obtained by a shooting method. The solution curves obtained by two
methods for the above chosen parameters are plotted in Figure 3.
\begin{figure}[H]
\begin{center}
\subfigure[ The parameter values are $vN=10^{-4}$ and $\chi=0.2$
($\alpha=4.95934$), $B=3A$ and $\lambda_0=1.077$. ]
{\includegraphics[width=2.4in]{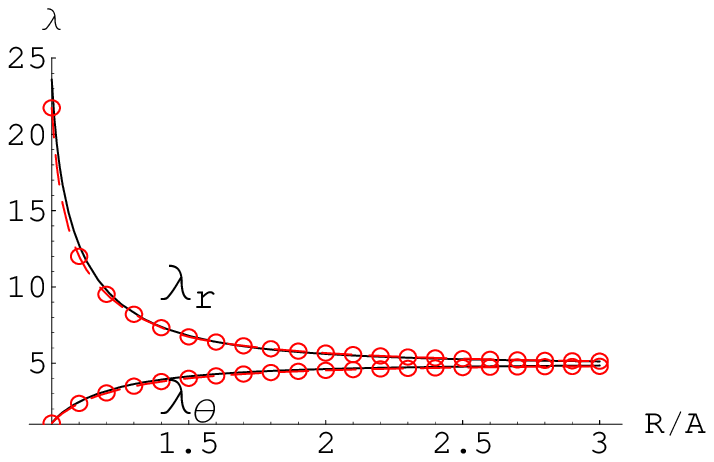}} \hspace{0.05in}
\subfigure[The parameter values are $vN=10^{-4}$ and $\chi=0.2$
($\alpha=4.95934$), $B=4A$ and $\lambda_0=1.077$.]
{\includegraphics[width=2.4in]{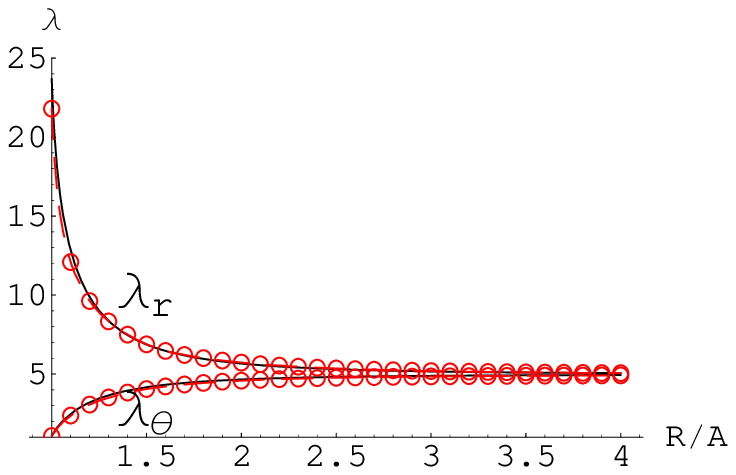}} \subfigure[ The parameter
values are $vN=10^{-4}\times2/3$ and $\chi=0.3$ ($\alpha=4.95934$),
$B=3A$ and $\lambda_0=1.077$. ]
{\includegraphics[width=2.4in]{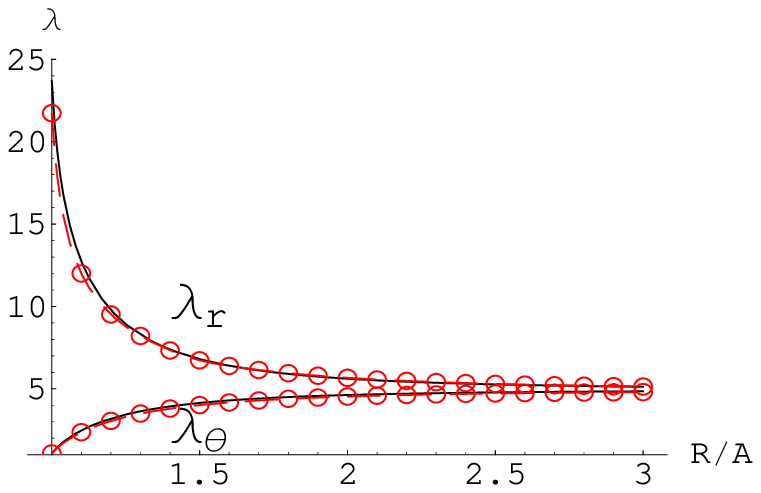}} \hspace{0.05in}
\subfigure[The parameter values are $vN=10^{-4}\times2/3$ and
$\chi=0.3$ ($\alpha=4.95934$), $B=4A$ and $\lambda_0=1.077$.]
{\includegraphics[width=2.4in]{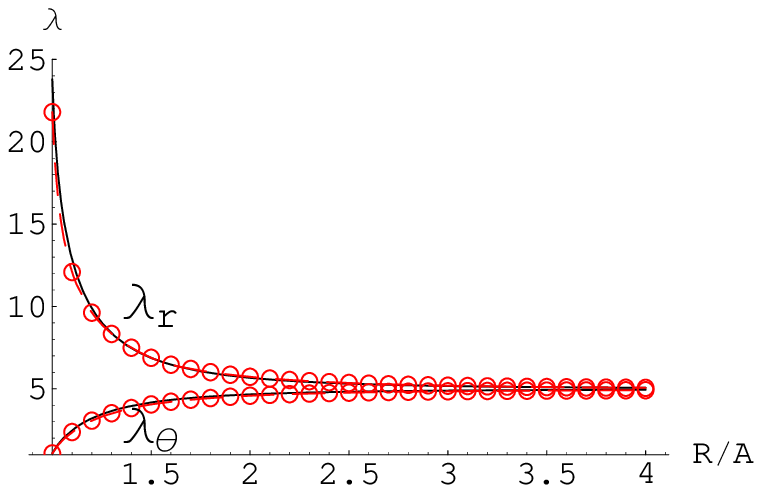}}
\end{center}
\caption{Distributions of the stretches (lines -o- are the analytical
solutions and black solid lines are the numerical solutions).}
\end{figure}

It can be seen that the analytical solution agrees well with the
numerical solution. Actually, the maximum relative errors of
$(\lambda_\theta,\lambda_r)$ are about $(2.7\%,7.7\%)$,
$(2.9\%,8.0\%)$, $(2.8\%,8.1\%)$, $(3.0\%,8.4\%)$ respectively for the
four figures. Keeping in mind that the obtained analytical solution is
only valid up to $O(1)$ and the $O(\epsilon)$ ($=O(\alpha^{-10/11})$)
terms are omitted, it produces reasonable good results already.

In Figures 3$(a,b)$ and Figures 3$(c,d)$, the values of $(vN, \chi)$
are different. However, in all cases the $\alpha$ value is same. Since
the analytical solution depends only on the $\alpha$ value, the
analytical curves in Figures $3(a,c)$ and Figures $3(b,d)$ are the
same respectively. The agreement between the analytical and numerical
solutions once again shows that the deformation is mainly determined
by the single material constant -- the hydrogel deformation constant
$\alpha$.

Now we shall give a more explicit expression than equation
$(\ref{eq49})$ for the inner solution. As it can be seen from
$(\ref{eq60})$ that $C_1$ is small, we seek a perturbation expansion
solution of equation $(\ref{eq48})$ of the form
\begin{equation}
\label{eq61} \hat{r}_{in}=r^*_0(X)+C_1 r^*_1(X)+\cdots.
\end{equation}
Substituting the above expansion into equation $(\ref{eq48})$ and
using the boundary condition $\hat{r}(0)=\hat\lambda_0$, we obtain
\begin{equation}
\label{eq62}
\begin{aligned}
\hat{r}_{in}(s)=&\left(\frac{5 \sqrt[3]{4}}{3} X+ \hat\lambda_0^{5/3}
\right)^ {3/5}+\\
&\frac{C_1}{{18 \sqrt[3]{2}}} \left[\left(\frac{5 \sqrt[3]{4}}{3}
X+\hat\lambda _0^{5/3}\right)^{7/5} -\hat\lambda_0^3\left(\frac{5
\sqrt[3]{4}}{3} X+\hat\lambda _0^{5/3}\right)^{-2/5} \right],
\end{aligned}
\end{equation}
where $X=s/(a \epsilon) = \alpha^{10/11}s/a$. Then $\tilde{r}_{in}$
can be immediately recovered by $\tilde{r}_{in}(s)=\alpha^{5/11}
\hat{r}_{in}(s)$. In Figure 4, we plot the solution curves of
$(\ref{eq49})$ and $(\ref{eq62})$. It can be seen that the difference
is very small. This supports the validity of the more explicit
expression $(\ref{eq62})$.
\begin{figure}[H]
\begin{center}
\subfigure[ The parameter values are $vN=10^{-4}$ and $\chi=0.2$,
$B=3A$ and $\lambda_0=1.077$. ]
{\includegraphics[width=2.4in]{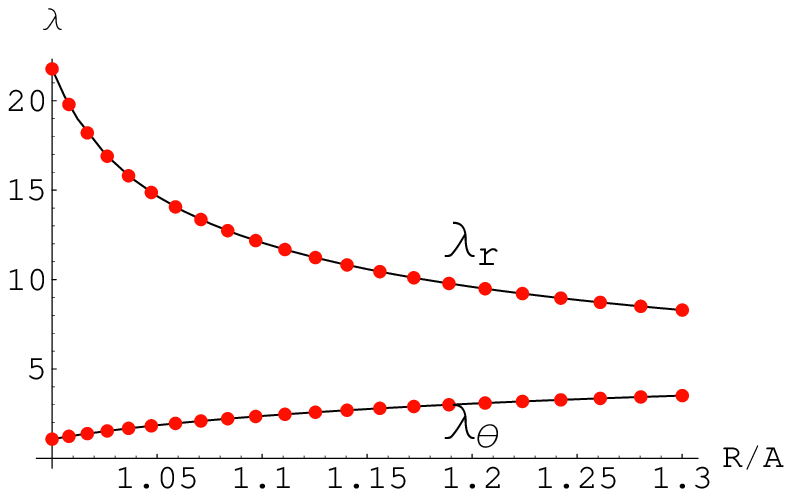}} \hspace{0.1in}
\subfigure[The parameter values are $vN=10^{-4}$ and $\chi=0.2$,
$B=4A$ and $\lambda_0=1.077$.]
{\includegraphics[width=2.4in]{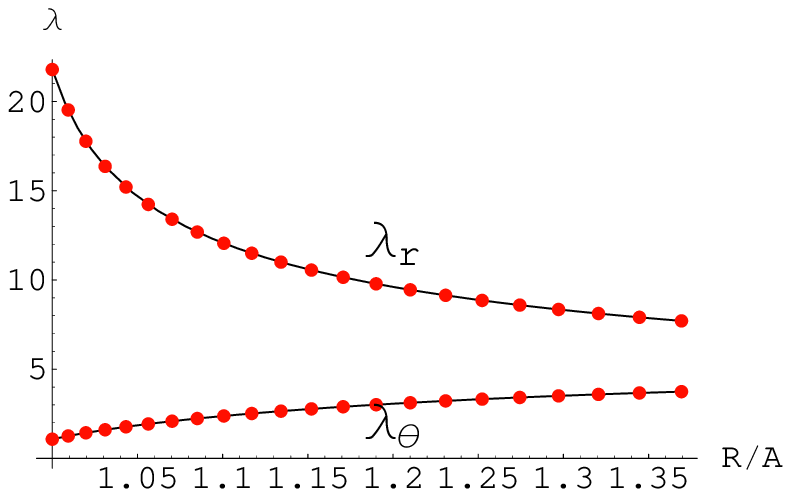}}
\end{center}
\caption{Comparison of the two inner solutions (dots are the solution
(\ref{eq49}) and solid lines are the explicit solution (\ref{eq62})).}
\end{figure}

Since we have obtained the analytical solution, some simple
approximate analytical formulas for important physical quantities can
be deduced. As in the previous section, we consider the stresses and
stretches at the inner and outer surfaces. At the inner surface $R=A$,
we have
\begin{equation}
\label{eq63}
\begin{array}{ll}
\displaystyle \lambda_\theta=\lambda_0 ,\ &\displaystyle
\lambda_r=\frac{\sqrt[3]{4}\alpha^{5/3}}{\lambda_0^{2/3}}+\frac{C_1
\alpha^{35/33}
\lambda_0^{2/3}}{3 \sqrt[3]{4} },\\
\displaystyle \frac{s_\theta}{NkT}= -\frac{\alpha^{10/3}}{\sqrt[3]{4}
\lambda_0^{7/3}} + \frac{ C_1 \alpha^{30/11}}{12\lambda_0}, \ &
\displaystyle \frac{s_r}{NkT}= \frac{3 \alpha^{5/3}}{2 \sqrt[3]{2}
\lambda_0^{2/3}}+ \frac{  C_1 \alpha^{35/33} \lambda_0^{2/3}}{2
\sqrt[3]{4} }.
\end{array}
\end{equation}
We can see that radial stretch $\lambda_r$ and radial stress $s_r/NkT$
are $O(\alpha^{5/3})$ quantities, while the circumferential stress
$s_\theta/NkT$ is much larger, an $O(\alpha^{10/3})$ quantity. What is
more, if $C_1$-terms are neglected ($C_1$ is small), all the stresses
and stretches at the inner surface are unaffected by the geometric
parameter $a$, the ratio of the inner radius to the thickness.  It
should be pointed out that, although $C_1$ is small, its value depends
on $a$. In the above cases, the $C_1$-terms in $(\ref{eq63})$ only
have a minor influence (less than $0.3\%$, compared with the first
terms).

At the outer surface $R=B$, we have
\begin{equation}
\label{eq64}
\begin{array}{ll}
\displaystyle \lambda_\theta=\alpha-\alpha^{-1} \frac{1-9C_2}{4},\quad
&\displaystyle \lambda_r=\alpha+\alpha^{-1} \frac{1-3C_2}{2},\\
\displaystyle \frac{s_\theta}{NkT}=\frac{-3(1-5C_2)}{2} \alpha^{-1},
\quad &\displaystyle \frac{s_r}{NkT}= 0.
\end{array}
\end{equation}
Comparing with equation $(\ref{eq10})$, we see that the stretch
$\lambda_\theta$ is close to but a little smaller than
$\lambda_{free}$ while the stretch $\lambda_r$ is close to but a
little larger than $\lambda_{free}$. The circumferential stress
$s_\theta/(NkT)$ is of $O(\alpha^{-1})$, which implies that a very
small stress needs to be applied to maintain this spherically
symmetric deformation.

If we impose the boundary condition $s_\theta(B)=0$ instead of
$s_r(B)=0$, the analytical solution can be constructed by the same
procedure described above. Actually, in this case only the
expression of the outer solution changes to
\begin{equation}
\label{eq65} \tilde{r}_{out}(s)=\alpha \frac{s+a}{a}+ \alpha^{-1}
\left[\frac{(1-5C_2)(1+a)^3}{2a(s+a)^2} + \frac{C_2(s+a)}{a} \right].
\end{equation}
The expressions of the inner and transition solutions and the
connection conditions are all the same. Also, the stresses and
stretches at the inner surface are still given by equation
$(\ref{eq63})$. And, at the outer surface we have
\begin{equation}
\label{eq66}
\begin{array}{ll}
\displaystyle \lambda_\theta=\alpha+\alpha^{-1}
\frac{1-3C_2}{2},\quad
&\displaystyle \lambda_r=\alpha+\alpha^{-1} (6C_2-1)\\
\displaystyle \frac{s_\theta}{NkT}=0, \quad &\displaystyle
\frac{s_r}{NkT}=3(5C_2-1)\alpha^{-1}.
\end{array}
\end{equation}
We see that in this case an $O(\alpha^{-1})$ tensile stress needs to
be applied at the outer surface to maintain this deformation.

\section{Conclusions}
We study analytically three cases of hydrogel swelling for a
core-shell structure, i.e. a free-swelling deformation, a near-free
swelling deformation and a general inhomogeneous deformation. The
hydrogel deformation constant $\alpha$, which is a combination of the
two material parameters $vN$ and $\chi$, is identified, and it is
found that this single material parameter plays a dominant role for
the deformations in all three cases. For the free swelling
deformation, a simple formula for the stretch $\lambda_{free}$ is
obtained in terms of $\alpha$ up to $O(\alpha^{-1})$. For the
near-free swelling one, we obtain the analytic solution for the whole
region. Some analytical formulas for the stresses and stretches at the
inner and outer surfaces are given, and it turns out they depend
linearly on the value $\lambda_0-\alpha$ (where $\lambda_0$ is the
given stretch at the inner surface). In this case, for a thick shell
(the ratio $a$ of the inner radius to the shell thickness is small),
the geometrical parameter $a$ has little effect. When the shell is
thin it is not stress-free in the circumferential direction at the
outer surface, which indicates the boundary conditions $s_r(B)=0$ and
$s_\theta(B)=0$ are not equivalent. For the general inhomogeneous one,
we treat it as a boundary layer problem. An extended method of matched
asymptotics is introduced to solve this problem. More specifically a
transition region is introduced to connect the inner region and outer
region. Further, we seek a series solution in the transition region
and impose proper connections with the inner and outer solutions.
Then, theoretically the problem is reduced to solve a system of 4
algebra equations. Analytical formulas for the radial and hoop
stresses and stretches at the inner surface are obtained. It is found
that both the radial and hoop stresses are very large and the former
is an $O(\alpha^{5/3})$ quantity while the latter is an
$O(\alpha^{10/3})$ quantity. Also, these quantities are independent of
the geometric parameter $a$ (to the leading order). Numerical
comparisons are also performed, and the results are in good agreement
with the analytical ones.

\section*{Acknowledgments}
The work described in this paper was supported by a grant (Project
No.: CityU101009) from the Research Grants Council of the HKSAR, China
and a strategic grant (Project No.: 7008111) from City University of
Hong Kong.



\begin{thebibliography}{00}

\bibitem{hydrogelballauff1} M. Ballauff and Y. Lu, ``Smart" nanoparticles:
preparation, characterization and applications, {\it Polymer} {\bf 48}
(2007) 1815--1823.

\bibitem{hydrogelbariffs1} A. Biffis and L. Minati, Efficient aerobic oxidation
of alcohols in water catalysed by microgel-stabilised metal
nanoclusters, {\it Journal of Catalysis} {\bf 236} (2005) 405--409.

\bibitem{hydrogelbariffs2} A. Biffis, N. Orlandi and B. Corain,
Microgel-Stabilized metal Nanoclusters: Size Control by Microgel
Nanomorphology, {\it Advanced Materials} {\bf 15} (2003) 1551--1555.

\bibitem{hydrogelbiot2} M. A. Biot, General theory of
three-dimensional consolidation, {\it Journal of Applied Physics} {\bf
12} (1941) 155--164.

\bibitem{hydrogelbiot1} M. A. Biot, Nonlinear and semilinear
Rheology of Porous Solids, {\it Journal of Geophysical Research} {\bf
78} (1973) 4924--4937.

\bibitem{perturbation2} A. W. Bush, {\it Perturbation methods for engineers and
scientists} (CRC Press, 1992).

\bibitem{hydrogelballauff2} J. J. Crassous and M. Ballauff, Imaging the Volume
Transition in Thermosensitive Core-Shell Particles by
Cryo-Transmission Electron Microscopy, {\it American Chemical Society}
{\bf 22} (2006) 6.

\bibitem{hydrogelkumacheva1} M. Das, S. Mardyani, W. C. W. Chan and E.
Kumacheva, Biofunctionalized pH-Responsive Microgels for Cancer Cell
Targeting: Rational Design, {\it Advanced Materials} {\bf 18} (2006)
80--83.

\bibitem{hydrogeldolbow1} F. Dolbow, E. Fried and H. Ji, Chemically induced
swelling of hydrogels, {\it Journal of the Mechanics and Physics of
Solids} {\bf 52} (2004) 51--84.

\bibitem{hydrogeldolbow2} F. Dolbow, E. Fried and H. Ji, A numerical strategy
for investigating the kinetic response of stimulus-responsive
hydrogels, {\it Computer Methods in Applied Mechanics and Engineering}
{\bf 194} (2005) 4447--4480.

\bibitem{hydrogeldurning1} C. J. Durning and K. N. Morman, Nonlinear
swelling of polymer gels, {\it Journal of Chemical Physics} {\bf 98}
(1993) 4275--4293.

\bibitem{hydrogelpolymer1} J. D. Ferry, {\it Viscoelastic Properties of Polymers}
(Wiley, 1980).

\bibitem{hydrogelpolymer2} P. J. Flory, {\it Principles of polymer chemistry}
(Cornell University Press, 1953).

\bibitem{hydrogelflory1} P. J. Flory and J. Rehner, Statistical
Mechanics of Cross-Linked Polymer Networks
\uppercase\expandafter{\romannumeral2}. Swelling, {\it The Journal of
Chemical Physics} {\bf 11} (1943) 521--526.

\bibitem{hydrogelgibbs1} J. W. Gibbs, On the equilibrium
of Heterogeneous Substances, {\it Transactions of the connecticut
Academy of Arts and Sciences} {\bf 3} (1878) 108--248.

\bibitem{hydrogelguo1} J. Guo, W. Yang, Y. Deng, C. Wang and S. Fu,
Organic-dye-coupted magnetic nanoparticles encaged inside
thermoresponsive PNIPAM microcapsutes, {\it small} {\bf 1} (2005)
737--743.

\bibitem{perturbation1} M. H. Holmes, {\it Introduction to perturbation methods}
(Springer-Verlag, 1995).

\bibitem{hydrogelsuo3} W. Hong, Z. Liu and Z. Suo, Inhomogeneous swelling of
a gel in equilibrium with a solvent and mechanical load, {\it
International Journal of Solids and Structures} {\bf 46} (2009)
3282--3289.

\bibitem{hydrogelsuo2} W. Hong, X. Zhao and J. Zhou and Z. Suo,
A theory of coupled diffusion and large deformation in polymeric gels,
{\it Journal of the Mechanics and Physics of Solids} {\bf 56} (2008)
1779--1793.

\bibitem{hydrogelkawaguchi1} H. Kawaguchi and K. Fujimoto,
Smart latexes for bioseparation, {\it Bioseparation} {\bf 7} (1999)
253--258.

\bibitem{hydrogellee1} J.-H. Kim and T. R. Lee, Discrete Thermally Responsive
Hydrogel-Coated Gold Nanoparticles for Use as Drug-Delivery Vehicles,
{\it Drug Development Research} {\bf 67} (2006) 61--69.

\bibitem{hydrogellyon1} S. Nayak, H. Lee, J. Chmielewski and L. A. Lyon,
Folate-Mediated Cell Targeting and Cytotoxicity Using Thermoresponsive
Microgels, {\it Journal of The American Chemical Society} {\bf 126}
(2004) 10258--10259.

\bibitem{hydrogelpence1} T. J. Pence and H. Tsai, Swelling-induced
microchannel formation in nonlinear elasticity, {\it IMA Journal of
Applied Mathematics} {\bf 70} (2005) 173--189.

\bibitem{hydrogelpence2} T. J. Pence and H. Tsai, On the cavitation
of a swollen compressible sphere in finite elasticity, {\it
International Journal of Non-Linear Mechanics} {\bf 40} (2005)
307--321.

\bibitem{hydrogelpence3} T. J. Pence and H. Tsai, Bulk Cavitation
and the Possibility of Localized Interface Deformation due to Surface
Layer Swelling, {\it Journal of Elasticity} {\bf 87} (2007) 161--185.

\bibitem{hydrogeltsai1} H. Tsai, T. J. Pence and E. Kirkinis, Swelling
Induced Finite Strain Flexure in a Rectangular Block of an Isotropic
Elastic Material, {\it Journal of Elasticity} {\bf 75} (2004) 69--89.

\bibitem{hydrogelsuo1} X. Zhao and W. Hong and Z. Suo, Inhomogeneous
and anisotropic equilibrium state of a swollen hydrogen containing a
hard core, {\it Applied Physics Letters} {\bf 92} (2008) 051904.

\end{thebibliography}

\end{document}